\documentclass[manuscript]{acmart}
\usepackage{csvsimple}
\usepackage{tabularx}
\newcolumntype{g}{X}
\newcolumntype{s}{>{\hsize=.5\hsize}X}

\usepackage{graphicx}
\usepackage{caption}
\usepackage{subcaption}
\usepackage{hyperref}
\usepackage{adjustbox}
\usepackage{todonotes}
\usepackage[compact]{titlesec}
\usepackage{multirow}
\usepackage{multicol}
\usepackage{booktabs}
\usepackage{tablefootnote}
\usepackage{bbm}

\usepackage[USenglish]{babel} 
\usepackage[nodayofweek,level]{datetime} 

\usepackage{algorithm2e}
\usepackage{algpseudocode}   
  
\usepackage{latexsym}
\usepackage{url}
\usepackage{float}

\AtBeginDocument{
  \providecommand\BibTeX{{
    \normalfont B\kern-0.5em{\scshape i\kern-0.25em b}\kern-0.8em\TeX}}}
    
\begin{document}

\title{Co-WIN: Really Winning? Analysing Inequity in India's Vaccination Response}

\author{Tanvi Karandikar}
\authornote{Authors contributed equally to this research.}
\affiliation{
  \institution{International Institute of Information Technology, Hyderabad}
  \country{India}
}
\email{tanvi.karandikar@students.iiit.ac.in}

\author{Avinash Prabhu}
\authornotemark[1]
\affiliation{
  \institution{International Institute of Information Technology, Hyderabad}
  \country{India}
}
\email{avinash.prabhu@students.iiit.ac.in}

\author{Mehul Mathur}
\authornotemark[1]
\affiliation{
  \institution{International Institute of Information Technology, Hyderabad}
  \country{India}
}
\email{mehul.mathur@students.iiit.ac.in}

\author{Megha Arora}
\affiliation{
  \institution{Palantir Technologies}
  \country{United Kingdom}
}
\email{marora@palantir.com}

\author{Hemank Lamba}
\affiliation{
  \institution {Dataminr, Inc.}
  \country{United States}
}
\email{hlamba@dataminr.com}

\author{Ponnurangam Kumaraguru}
\affiliation{
  \institution{International Institute of Information Technology, Hyderabad}
  \country{India}
}
\email{pk.guru@iiit.ac.in}

\newcommand{\COVID}{\textsc{COVID-19 }}
\newcommand{\cowin}{CoWIN dashboard }
\newcommand{\TK}[1]{{\color{purple}{\textbf{TK}: #1}}}
\newcommand{\AP}[1]{{\color{blue}{\textbf{AP}: #1}}}
\newcommand{\MA}[1]{{\color{orange}{\textbf{MA}: #1}}}
\newcommand{\RR}[1]{{\color{green}{\textbf{Reviewer}: #1}}}
\newcommand{\HL}[2]{{\color{red}{\textbf{HL}: #1}}}
\newcommand{\MM}[2]{{\color{red}{\textbf{MM}: #1}}}


\begin{abstract}
The \COVID pandemic has so far accounted for reported $5.5M$ deaths worldwide, with $8.7\%$ of these coming from India. The pandemic exacerbated the weakness of the Indian healthcare system. As of \formatdate{20}{1}{2022}, India is the second worst affected country with $38.2 M$ reported cases and $487K$ deaths. According to epidemiologists, vaccines are an essential tool to prevent the spread of the pandemic. India's vaccination drive began on \formatdate{16}{1}{2021} with governmental policies being introduced to prioritize different populations of the society. 
Through the course of the vaccination drive, multiple new policies were also introduced to ensure that vaccines are readily available and vaccination coverage is increased. However, at the same time, some of the government policies introduced led to unintended inequities in the populations being targeted. In this report, we enumerate and analyze the inequities that existed in India's vaccination policy drive, and also compute the effect of the new policies that were introduced. We analyze these potential inequities not only qualitatively but also quantitatively by leveraging the data that was made available through the government portals. Specifically, (a) we discover inequities that might exist in the policies, (b) we quantify the effect of new policies introduced to increase vaccination coverage, and (c) we also point the data discrepancies that exist across different data sources.

\end{abstract}




\maketitle

\section{Introduction}

\begin{figure}
   \centering
    \begin{subfigure}[b]{0.45\textwidth}
        \centering
    \includegraphics[width=1\linewidth]{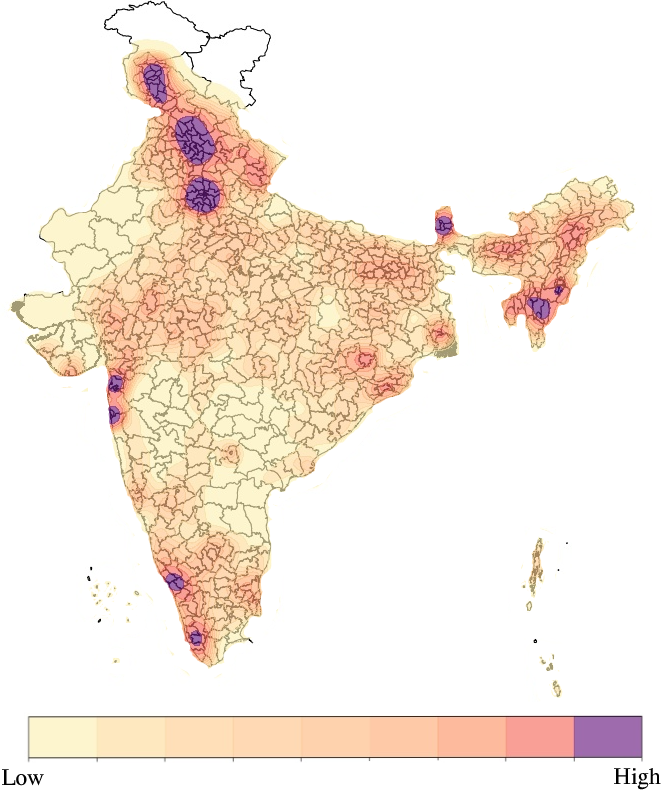}
 \end{subfigure}
    \begin{subfigure}[b]{0.5\textwidth}
        \centering
    \includegraphics[width=1\linewidth]{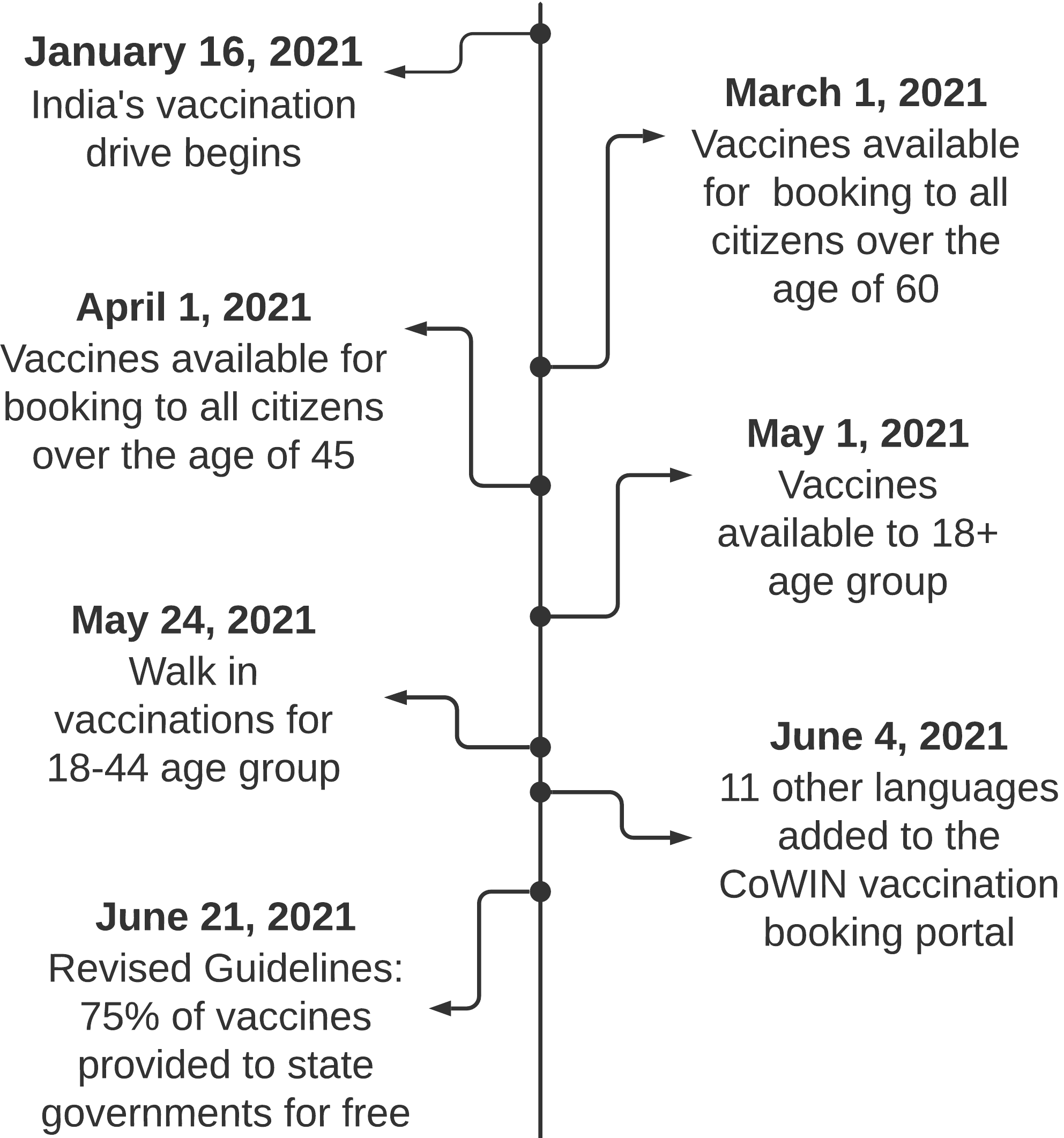}
\end{subfigure}
\caption{(\textbf{L}eft) Distribution of vaccines administered normalised by population in India, as of \formatdate{22}{7}{2021}. It is evident that a few areas have high density of vaccinations, while others seem to be very sparsely vaccinated. (\textbf{R}ight) Timeline of important COVID policies and phases in India.}
\label{fig:timeline}
\end{figure}

The \COVID pandemic has had a devastating effect worldwide, resulting in $345,975,189$ cases and $5,584,189$ deaths as of \formatdate{21}{1}{2022}~\cite{john_hopkins}.
Though the numbers have been high overall, some sections of society have been disproportionately affected by the pandemic. For example, in the US, African American and Latino populations have higher mortality rates than other races~\cite{Gross2020}. Similarly, Shadmi et al. note that the pandemic has disproportionately affected the poor, minorities, and more vulnerable sub-populations in terms of transmission, economic problems, and various other societal effects~\cite{Shadmi2020}. A report by the US Center for Disease Control states that sections of society with essential occupations like healthcare workers, grocery store workers, and individuals with lower income and lack of access to high-quality education and wealth are at an increased risk of getting the disease~\cite{race_ethnicity}. 

With the virus already aggravating existing inequalities, it is essential to ensure an equitable distribution of vaccines. The Director-General of the World Health Organisation said -- “Vaccine inequity is the world’s biggest obstacle to ending this pandemic and recovering from COVID-19”~\cite{undermine_global}. Most manufactured vaccines are supplied to affluent countries, with poorer nations struggling to vaccinate their population. According to the Global Dashboard for COVID-19 Vaccine Equity~\cite{undp_vaccine} (a joint initiative from UNDP, WHO and the University of Oxford’s Blavatnik School of Government), as of \formatdate{12}{11}{2021}, 64.99\% of the population has been vaccinated with at least one dose in high-income countries, while the same number stands at 6.48\% for low-income countries. The inequity in the availability of vaccines not only exists at a global scale, but also exists within each country -- many nations are struggling to make the distribution of vaccines equitable ~\cite{usa_inequity}. The non-availability of vaccines to marginalized populations further increases the uneven impact of the \COVID pandemic. For example, in the US, many black and brown communities are characterized by high housing density and poor access to healthcare services~\cite{inequity_covid_lowincome}. Additionally, members from these communities are employed in jobs where they cannot safely socially distance or do not have the privilege to work from home~\cite{yancy2020covid}. This phenomenon is not limited to developed countries and perhaps unravels more intensely in developing countries~\cite{Shadmi2020}. Researchers have mainly focused on showing how the impacts of \COVID are disproportionately affecting marginalized sections of society across every country in the world~\cite{inequity_covid_lowincome, Shadmi2020, Gross2020, systemic_inequity_us}. Every government has tried to introduce \COVID specific policies to mitigate these effects~\cite{hale2021global}. Some of these policies have been studied for potential inequities that they might have overlooked or introduced~\cite{eviction_mitigation,carly2021covid,grogan2021unsanitized}. However, most of these analyses are from a western or developed world perspective.\footnote{In this paper, we use the terminology of developed world and developing world, however they could be replaced by high-income or low-income respectively according to taxonomy used by World Bank.} Very limited work~\cite{time_series, bias_maharashtra} has been done to understand how \COVID mitigation policies have been inequitable, or how effectively they address disproportionate effects in the developing world. Moreover, the vulnerable population in developing countries might not follow the same definitions as that of vulnerable subsections in the developing world. 

In this paper, we focus our analysis on India because - (a) India is one of the worst countries to be hit by the pandemic, (b) Being the second-largest country by population, India's vaccination drive will be one of the largest vaccination drives, (c) India carries very different nuances for vulnerable populations and societal structure in comparison to most of the western world, and (d) The Government of India has made the data related to vaccinations available via CoWIN\footnote{https://www.cowin.gov.in/}, a web portal for COVID-19 vaccination registration, owned and operated by India's Ministry of Health and Family Welfare. Additionally, the implications of addressing inequitable policies can have a huge impact for India -- as of \formatdate{20}{1}{2022}, only 48.1\% of the Indian population is fully vaccinated ~\cite{dashboard_cowin}, in comparison to US (63.1\%)~\cite{cdc_percentage_us} and UK (70.2\%)~\cite{uk_percentage}.
Interestingly, as of June 4, 2021, the same number of vaccines had been administered in 114 of the least developed districts in India as were administered in 9 urban cities, which combined have half the population of the former~\cite{inequity_countryside}. This disparity is also evident in Figure~\ref{fig:timeline}(L) which shows the distribution of administered vaccines nation-wide normalised by population. With half of the population yet to be vaccinated, and optimal resource allocation still needed, we believe our work can provide actionables for the government to deal with during the coming phases of the drive and booster doses.

Past work analyzing the vaccination drive in India is limited in scope by focusing on particular states or policies and does not explicity address nationwide equitable distribution~\cite{bias_maharashtra, time_series, Foy2021, Babu2021}. In this paper, we concentrate on a more holistic perspective across the country while accounting for equity and fairness considerations. 
Specifically, we use the data made available by the government through the \cowin to answer the following questions:

\vspace{-0.75em}
\begin{enumerate}
    \item{\textbf{Inequities in Vaccination Drive:} What are the biases, if any, present in how vaccines have been made accessible and administered in India?}
    \item{\textbf{Effectiveness of vaccination policies:} How effective were the government policies in the vaccination drive towards increasing vaccination response and mitigating inequities?}
    \item{\textbf{Consistency of Data Sources:} How consistent are the different data sources made available by the government in terms of numbers being published to aid decision making?}
\end{enumerate}
\vspace{-0.75em}

We discovered that number of vaccination centres and even the total vaccinated percentage (Fig \ref{fig:timeline}) is not equitable across different sensitive attributes. Districts with more urban people had a higher percentage of vaccines administered to their population than the ones with a rural population in the majority. Throughout the vaccination drive, the government of India implemented many policies to ensure that vaccinations are made available to as many people as possible. We analyzed the effectiveness of two policies centred around making vaccination drive equitable by performing a Regression discontinuity analysis and discovered that both of the policies had a limited positive impact. For ensuring transparent and effective policymaking, it is critical that the data reported about the vaccination drive is accurate and granular. In Section~\ref{sec:preliminary}, we identify inconsistencies within the data sources published by the government. We discovered that the CoWIN dashboard API provides the number of vaccines administered, but the published figures do not add up to consistent totals, indicating significant discrepancies (Section~\ref{sec:dash_dis}).

\section{Related Work}
Our related work flows from three different directions - (a) studies related to inequity in how COVID affected minority populations, (b) studies related to inequity in or due to specific pandemic policies, and (c) quality of vaccination data.

\textbf{Inequity in COVID impact}: Previous works have highlighted the disproportionate impact of \COVID on disadvantaged sections of society~\cite{Wang2020inequity}. Studies have shown that the loss of jobs due to \COVID was more among persons of colour, women and immigrants~\cite{saenz2020jobinequity}. The rates of depressive symptoms were also higher among people of lower income groups~\cite{Ruth2020mentalhealth}. At the governing level, Dzigbede et al. show that poorer governments will not be able to respond well to the virus and social inequalities will grow~\cite{Dzigbede2020disaster}. Certain marginalized populations in the US such as African Americans, Latino individuals, and Native Americans experience an inordinate amount of COVID-19 related infections and deaths~\cite{Berkowitz2020inequity, Garcia2022systematicracism, systemic_inequity_us}. In such areas,  Social Determinants of Health (SDOH) like limited education, unemployment and structural racism contribute to health conditions like cancer and cardiovascular diseases which unfairly predisposes them to worse outcomes upon acquiring COVID-19~\cite{systemic_inequity_us}. Similar work focusing on role of racial, economic and ethnic inequities in the imbalanced vulnerability of populations in the US~\cite{racial_inequity_us} has been done by Lee et al. They used a social-ecological model which predicted 280,000 deaths if the country opened up fully or a loss of 18 million jobs in the case of a country-wide lockdown, mainly affecting the minority groups. \COVID outbreaks in jails and prisons are also  higher compared to other areas~\cite{FRANCOPAREDES2021e11}.


\textbf{Inequity in COVID mitigation policies}: Policies implemented during the pandemic play a critical role in alleviating the burden of \COVID on different sections of society. In China, when the government announced that \COVID treatments would be subsidized, it enabled many poorer citizens to get treated immediately, while even reducing the risk of transmission to others~\cite{Wang2020CombatingCH}. On the other hand, improperly implemented policies can further burden underserved sections of society. Glover et al. found that several policies around the world resulted in harms such as anxiety, depression, food insecurity, loneliness, stigma, violence~\cite{GLOVER202035}. In US, the "CARES Act of 2020" (which is supposed to provide relief funds to hospitals) provided more funds to hospitals that cater to privileged sections of society as opposed to hospitals that cater to low-income people of color~\cite{Colleen2021bailout}. Benfer et al.~\cite{eviction_mitigation} studied how eviction and uncertainty in housing policies during the pandemic subverted COVID-19 mitigation policies.
Eviction increases the risk of contracting COVID-19 due to crowding of people and bad living conditions especially in the neglected and economically underdeveloped areas, thus making mitigation policies such as social distancing impractical to implement. 
Emmanuel et al.~\cite{migrants_slums_inequality} used the Indian context and examined the effect of a lockdown and social distancing government guidelines on urban slums and migrant workers. The sudden and unplanned nature of the lockdown caused migrant workers to be unemployed and stranded in their current working locations, unable to return to their villages. They also highlight how government guidelines such as social distancing are impossible to follow in informal settlements such as slums (where the migrant workers reside) which have high population density. Thus previous work has highlighted how government policies regarding handling of the pandemic were not equitable towards all sections of society. In Section \ref{sec:macro}, we further examine if the government policies pertaining to vaccination have been equitable.


\textbf{Data quality analyses}: While not much work has been done on the quality of vaccination data, previous works have looked at the quality of other reported statistics regarding COVID-19 in India (number of cases, deaths reported, resource availability, etc.). Vasudevan et al.~\cite{data_sources} studied the quality of reporting of COVID-19 data in over one hundred government platforms from India. Vasudevan et al. found lack of granularity in the reporting of COVID-19 cases, vaccinations and vacant bed availability. They also found that age, gender and comorbidity was available for less than 30\% of cases and deaths. There was no reporting of adverse events following immunization by vaccine and event type. Zimmermann et al.~\cite{cases_underreporting} found that the number of cases in India was under-reported by a factor of 10 to 20, and the number of deaths were under-reported by a factor of 2 to 5. Similarly Jha et al.~\cite{under_reported_mortality} studied and collected data about \COVID mortality in India. They found that the number of actual deaths from \COVID in India is significantly greater than that of deaths reported officially. Using a representative and distributed study (survey), they reported that the number of collective deaths till September 2021 is 6 to 7 times greater than the number in official reports. More importantly, most of the work centered around the \COVID vaccination drive does not mention or take into account any inequities the policies might have introduced. To the best of our knowledge, no previous work has studied the consistency between the various data sources made available by the government on the vaccination drive.

\textbf{Inequity in vaccine distribution}: The inequitable distribution of vaccines on a global scale was studied by Moosa et al.~\cite{global_inequality}. They used Lorenz Curves and Gini Coefficients~\cite{lorenz_gini} to illustrate the unequal distribution of vaccine stockpiles due to hoarding by the wealthier countries. They found that 80\% of the population had only 5\% of the total COVID-19 vaccines in the world. The value of the Gini coefficient, which ranges from 0 to 1 with 1 representing perfect unequal distribution, was found to be 0.88 for COVID-19 vaccines around the world. Bolcato et al. analyze how the worldwide distribution of vaccines disregards the principle of equity~\cite{equitable_vaccines}. To the best of our knowledge, no work has been done to study the inequity of vaccine distribution in India, and we aim to bridge this gap in our work.






\section{Data Sources}
 
The government authorities made data available related to \COVID vaccinations through 2 different APIs -- (a) CoWIN Dashboard API and (b) CoWIN Portal API. We review each of these APIs and then describe the data we leverage for our quantitative analysis. 

\subsection{CoWIN dashboard APIs}
\label{sec:cowin_dashboard_API}
The Government of India reports the number of vaccines administered in each district via the CoWIN dashboard, updating the dashboard every few minutes~\cite{dashboard_cowin}. Through the underlying APIs, we are able to obtain information about the number of vaccines administered and the different sub-groups (for example -- based on age, gender) to which vaccines were given. This is a rich data source and we rely on it for most of our quantitative analysis. Though the data from the dashboard is relatively consistent, we did identify certain discrepancies that exist in the data, which we report in Section~\ref{sec:dash_dis}. One limitation of this data source is that sub-groups are not present for all past dates and districts~\cite{data_sources}.

\subsection{Census Data}
\label{sec:census}
To evaluate equity of vaccine distribution among different sections of society, we require demographic data at an all-India level. For this purpose, we use the most recent Census of India~\cite{census_india}, which was conducted in 2011. The Census of India provides data for every district in every state. We choose district as the granularity level for our analysis analysis largely because the vaccination data from the dashboard is only available at the district level. 

According to census, urban area is defined as a town which has (a) minimum population of 5,000, (b) at least 75 percent of the male main workers are engaged in non-agricultural pursuits; and (c) density of population of at least 400 per sq. km. However, this classification only holds for census towns. To be able to aggregate rural-ness or urban-ness of a district, we defined \emph{rural ratio} as the fraction of population living in rural areas in that district. A higher rural ratio for a particular district indicates higher fraction of people living in rural areas.

Additionally, the protected variables (i.e. often known as \emph{sensitive attributes} in machine learning literature) vary drastically in Indian context. For instance, besides gender, age and urbanicity (indicated by rural-ness or urban-ness of a district), Indian society also has inequities related to religion, caste or literacy levels~\cite{sambasivan2021re}. Indian census contains detailed information about the attributes relevant to these factors at district level - and we leverage them to conduct our analysis. Specifically, we leverage the percentage of SC (Scheduled Caste) and ST (Scheduled Tribe) related information. Both of these groups are officially designated by Indian government, and widely considered as the most disadvantaged socio-economic groups. They account for roughly 16\% and 8\% of the total Indian population, according to Census 2011.



Because the naming and division of districts within each state has changed since the last census, we formed a mapping of old districts to new, merging the districts where borders had been changed to form "pseudo districts". We adjusted demographic values by aggregating values of the districts being merged.  
We are open-sourcing this mapping of old to new districts. \footnote{\url{https://zenodo.org/record/5889637}}



\section{Distribution of Administered Vaccines and Centers}
\label{sec:correlation}

Access to healthcare has been known to be highly disparate towards various subsections of the population~\cite{Riley2012, Weiss2020}. Researchers have discovered significant differences among different socio-economic groups regarding accessibility to healthcare. This disparity exists between different countries (number of hospital beds in low income countries as 9 per 10K in comparison to 57 for high income countries~\cite{peters2008poverty}) as well as within country e.g. for India, the access to healthcare and expenditure by respective governments on the citizens was significantly different between urban and rural states~\cite{balarajan2011health}. We believe that even during COVID-19, access to healthcare could have been a crucial factor hindering the equity of vaccinations to the population. Applying the same intuition to our work, we believe that access to vaccination health centers and consequently the number of vaccines that were administered will exaggerate the inequities.




To conduct this analysis, we obtain vital statistics i.e., the number of healthcare centers in each district and the number of vaccines administered by each center through the CoWIN Dashboard API. Both of these metrics are indicative of the magnitude of presence of vaccination drive in that district. The number of health centers metric gives an insight into how many dedicated centers were set up, and if people had accessible options to get their vaccination. This can also speak to waiting times -- as a lower number of health centers after normalizing for population can lead to higher waiting times and a higher fallout rate.
Additionally, more number of centers (assuming population density is uniformly distributed) could also mean easy access to healthcare centers based on average commute time or average waiting time. Similarly, the number of vaccines administered in each district provides a ground reality indicator for what fraction of the population was vaccinated in that district.

For using the above mentioned variables for any statistical analysis, we normalize them by the population of the district (per 100K individuals). Borrowing the hypothesis from past work, we study the above mentioned vaccine accessibility and vaccine coverage variables along with the rural-ness and urban-ness of the district, the literacy rate, gender ratio and the ratio of Scheduled Caste and Scheduled Tribe population. To quantify the rural-ness of a particular district, we use \emph{rural ratio} (as defined in Section~\ref{sec:census}). The vaccines and centers data for this analysis was collected from CoWIN dashboard on \formatdate{22}{7}{2021} and \formatdate{22}{10}{2021} (i.e. 82 days and 174 days respectively after the vaccination drive opened for all citizens in 18+ age group).

\subsection{Preliminary Analysis}
\label{sec:pearsons}
\begin{figure*} \centering
  \begin{subfigure}[b]{0.45\columnwidth}
  \centering
     \includegraphics[width=\linewidth]{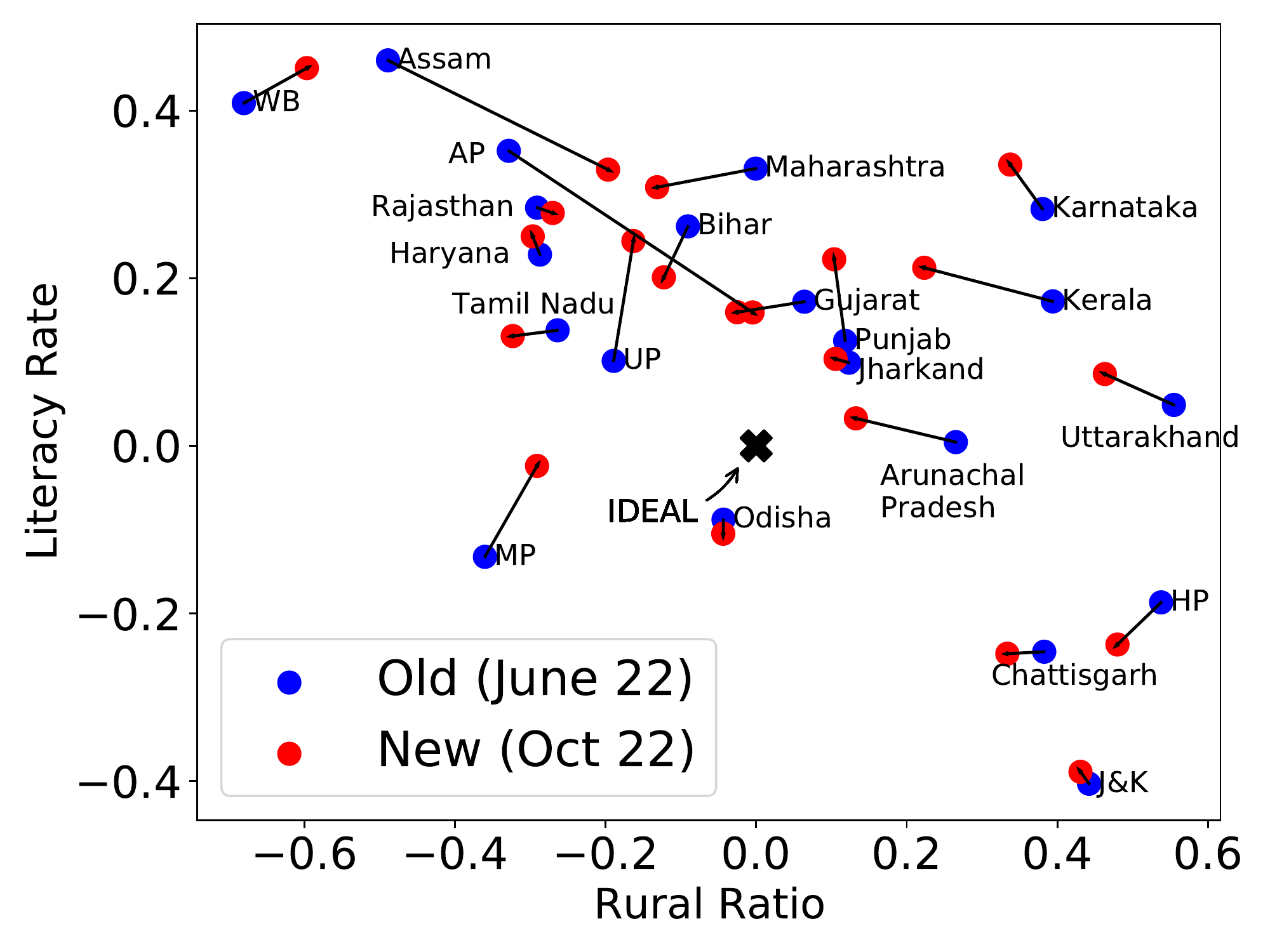}
     \label{Fig:vaccinations_API}
  \end{subfigure}
  \begin{subfigure}[b]{0.45\columnwidth}
     \centering
     \includegraphics[width=\linewidth]{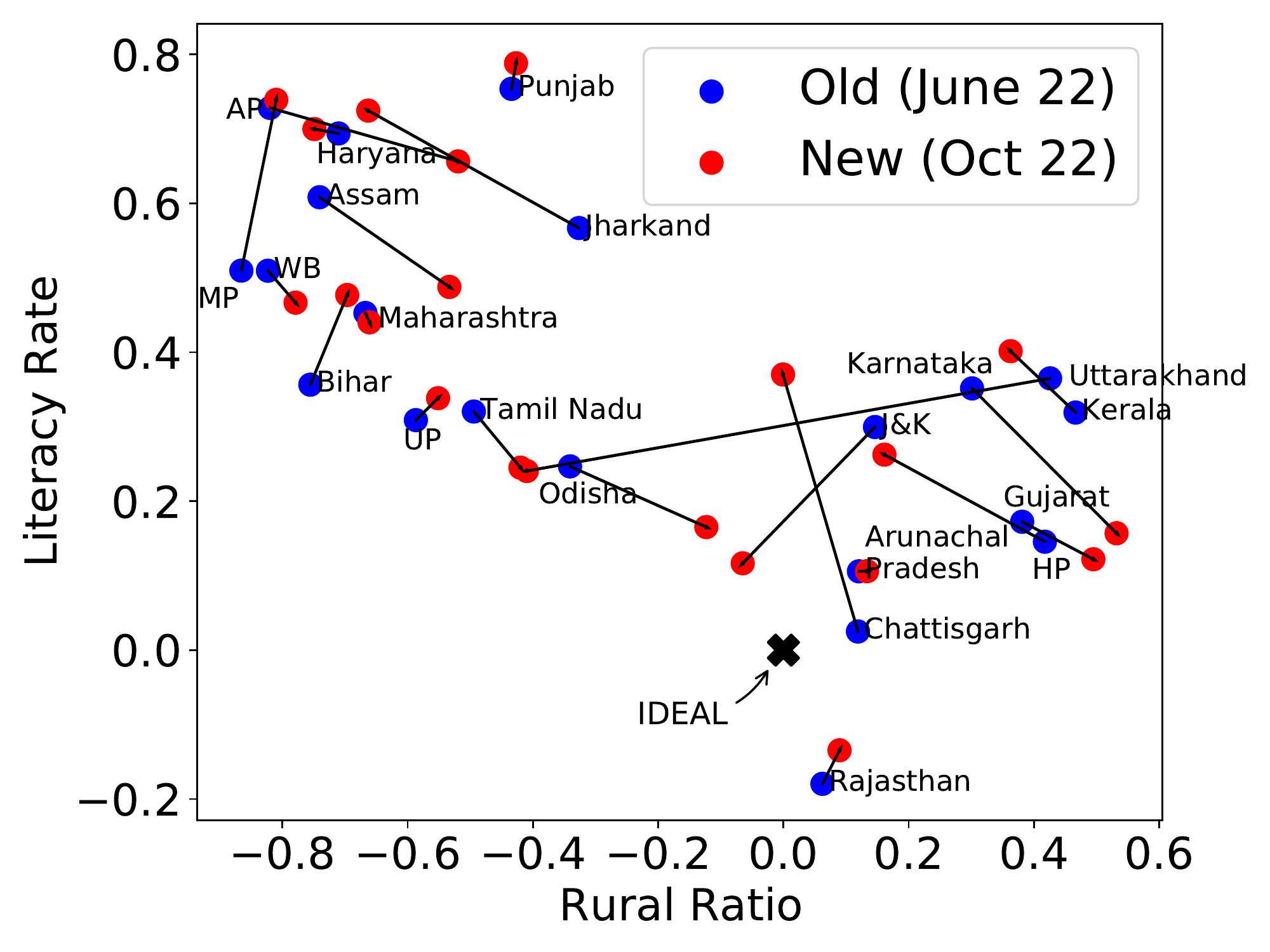}
     \label{Fig:vaccinations_sub}
  \end{subfigure}
\caption{
Pearson Correlation computed for every state, at district level between Rural Ratio (X-axis), Literacy Rate (Y-axis) and (\textbf{L}eft) Number of Vaccination Centers per 100K and (\textbf{R}ight) Total Vaccination percentage. The blue points show the correlation on June 22nd, and red points show correlation 5 months later on October 22nd. We can observe that most states have negative correlation with rural ratio, i.e., as rural ratio increases, the number of vaccination centers or the total vaccination percentage decrease. Similarly, we see most states have positive correlation with literacy rate i.e. as literacy rates increase, the number of vaccination centers or the total vaccination percentage increases}
  \label{fig:correlation}
\end{figure*}

For preliminary analysis, we rely on Pearson correlation to identify how states are performing in terms of (a) making available vaccination centers and (b) total vaccination percentage. We correlate the performance of each state on these target variables along sensitive societal variables - (i) rural ratio and (ii) literacy rate. A state performing well, i.e., being equitable along these sensitive variables will have no or low correlation between performance and sensitive variables. We show the performance of these states in Figure~\ref{fig:correlation}. We can observe that most states have negative correlation with rural ratio, i.e., as rural ratio increases, the number of vaccination centers or the total vaccination percentage decrease.  Similarly, we see most states have positive correlation with literacy rate i.e. as literacy rates increase, the number of vaccination centers or the total vaccination percentage increases. Even when noticing temporal trends i.e. between two dates three months apart, we note that most states try to improve on the performance metrics and try to move closer to the equitable line.

\subsection{Spatial Regression}
We rely on spatial regression models to model the relationship between districts and their relative performance on the number of vaccination centers made
available and the total percentage of vaccinated individuals. Spatial regression models allow to identify the trend while controlling for spatial lag effects. The spatial lag is the spillover of the value of the feature on the surrounding districts (i.e., change in value for a particular district for unit change in its neighbour's value). Each district's neighbouring districts are found using the Queen-move algorithm~\cite{spatial_weights}. Overall the spatial regression model can be written as:
\begin{equation}
    \label{eqn:spatial_lag_vector}
    Y = \alpha + X\beta + WX\lambda + \epsilon
\end{equation}
where $Y$ is the target variable, $X$ is the set of covariates, $W$ captures the spatial weights, and as in regression, $\beta$ and $\lambda$ capture the effect; $\alpha$ is the intercept and $\epsilon$ captures the residuals.

We use target variables as the number of vaccination centers per district, and log of total number of vaccines. To understand the effect between different sensitive variables and the performance variables, we include multiple sensitive attributes (after normalizing for population) as mentioned above. To control for the population variance that might exist in the data, we include control for the logarithm of population. Similarly, to account for the vastness of the district, we include the logarithm of area as well.

The results of the regression analysis are shown in Table \ref{tab:vac_regression}. We observe that in the case of vaccination centers as the target variable, every feature except literacy rate and Scheduled Tribe (ST) population ratio is significant. Rural ratio and Scheduled Caste (SC) ratio have a significant negative coefficient, i.e., districts with higher rural ratio and SC population tend to have lower number of centers meaning that there is some inequity based on these two features. For vaccines as target variable, the ratios of Rural, SC and ST populations all have a significant negative effect. In this case, literacy rate has a notable positive effect on the number of vaccines made available. The difference in how literacy rate affects one response variable but not the other is intriguing. We hypothesize that this could be due to the fact that vaccine booking and administration is something that is more affected by the characteristics of the general population, unlike decisions around vaccination center planning, which are made by the government. Area, as expected, has a high significant positive effect on the number of centers.


\begin{table}[!hbtp]
\caption{Spatial regression on number of centers and number of vaccines administered in each district.}
\begin{tabular}{ccccc}
\toprule
Features & \multicolumn{2}{c}{\# Vaccination Centers} & \multicolumn{2}{c}
{$\log(\text{Vaccinations})$} \\ 
\cline{2-5}
 & Coefficients (SE) & LR ChiSq & Coefficients (SE) & LR ChiSq  \\
\midrule
 $\log$(Population) & $339.09(28.16)^{***}$ & 131.54 
 & $0.768(0.03)^{***}$ & 360.84 \\
 $\log$(Area) & $159.81(28.18)^{***}$ & 31.74 
 & $0.108(0.03)^{***}$ & 9.68 \\
 Male Ratio & $1973.41(637.57)^{***}$ & 9.63 
 & 0.58(0.78) & 0.55 \\
 Rural Ratio & $-387.43(66.46)^{***}$ & 33.49 
 & $-0.31(0.08)^{***}$ & 14.76 \\
 Literacy Rate & $97.58(117.61)$ & 0.69 
 & $0.57(0.14)^{***}$ & 15.65 \\
 SC Ratio & $-450.31(129.10)^{***}$ & 12.20 
 & $-0.8(0.15)^{***}$ & 25.10 \\
 ST Ratio & $-60.79(49.51)$ & 1.52 
 & $-0.13(0.05)^{**}$ & 4.97 \\
\midrule
$R^{2}$ coefficient & \multicolumn{2}{c}{0.43} & \multicolumn{2}{c}{0.63}\\
\bottomrule
\multicolumn{5}{c}{Note:$^{***}p<0.001$, $^{**}p<0.01$, $^{.}p<0.1$}
\end{tabular}

\label{tab:vac_regression}
\end{table}

\section{Inequities in Government Policies}

\label{sec:macro}

It is not a point to be re-emphasized, but the government's response in handling a pandemic is crucial, and no effective intervention through policies or laws can affect health and well-being adversely~\cite{OECD2020}. Specifically, effective implementation of government policies can decrease the rate of spread, limit the number of hospitalizations and deaths~\cite{Fuller2021}. However, it is possible that many of the proposed policies could be inequitable towards a certain section of the society which might lead to amplification of existing societal (especially healthcare-related) inequities~\cite{eviction_mitigation, migrants_slums_inequality}. In this section, we enumerate and analyze some of the major policy decisions related to vaccination drive implemented by the government of India at the national level from an equity perspective.

From January 16th (the day vaccine drive was launched) to May 24th (walk-in vaccinations were made available) (see Figure~\ref{fig:timeline}),  the only way of getting vaccines was to obtain a vaccine slot using CoWIN app / website.  Even after May 24th, CoWIN app/website remained the primary way of getting vaccines and providing citizens with certificate proof of administered vaccinations~\cite{cowin_sole_platform}. The primary reliance on a technological solution, which faced issues related to latency~\cite{cowin_glitch2} and getting vaccination appointments being challenging~\cite{vacc_appts_challenging}, led to inequity in vaccination drive from the beginning.

\subsection{Barriers to accessing CoWIN app}
\label{sec:cowin_rural}



It is well known that access to health centers is inequitable~\cite{Riley2012, Weiss2020} across the world, however in the case of India's vaccination drive, the potential inequities started seeping through even before the patient is asked to get to the health center. The primary way of getting vaccine appointments i.e. CoWIN app is primarily a web portal for vaccine appointment booking, which requires access to the Internet. Even with 100\% increase in number of people with access to Internet in the past 5 years, the number of internet subscribers in India as of March 2021 stands at $825.30M$, which accounts for roughly $58\%$ of the population~\cite{trai}. The reliance on access to Internet puts a large portion of the Indian population at a disadvantage of not getting vaccines in a timely manner. Moreover, the demographics of people who do not have access to the Internet are not uniform and centered on the subsections of people who are already societally disadvantaged. According to the IAMAI-Kantar ICUBE 2020 report ~\cite{kantar_icube}, 67\% of the urban population and 31\% of the rural population are active Internet users. This digital divide is present across genders as well; 58\% of the active Internet users in India are men, and the remaining are women (the report does not mention anything about non-binary genders). Given that India's rural population stands at 66\% ~\cite{rural_urban_density}, the introduction of an Internet-based registration system has the potential to exacerbate these existing inequities.


\label{subsec:language}
Since its release to the general public on \formatdate{1}{3}{2021} up until \formatdate{4}{6}{2021}, CoWIN was only available in English. According to the 2011 census data, only 10.67\% of Indians speak English as their first, second or third language~\cite{english_cowin_bias}. 
According to a Lok Foundation Survey~\cite{english_urban_rich}, English speakers are richer, more educated, and likely belong to
upper castes.  There is also a religious disparity -- over 15\% of Christians can speak English, as opposed to less than 6\% of Hindus and 4\% of Muslims. Only about 3\% of the rural population could speak English as opposed to the 12\% in the urban population. For these reasons, the most vulnerable were often those who could not book vaccination appointments due to comprehensibility issues. For example, daily wage workers who were least likely to have the knowledge and means to use a smartphone application did not have any incentives to forgo the day’s wages to get vaccinated~\cite{apnews}, while being the ones who did not have the privilege to work remotely.


\subsection{Impact of releasing the CoWIN API}
\label{sec:cowin_api_bot}


The CoWIN API released by the government makes the following functionalities available: (a) Discover vaccination centers and related information, (b) Schedule appointments, (c) Manage the scheduling and logging of vaccination administration (mainly for \COVID Vaccination Centers), (d) Generate/download certificates, (e) Report any adverse events after vaccination as per AEFI (Adverse Events Following Immunization) guidelines. These API services were released with the intention to be advantageous to the population. However, not all sections of the population could reap these benefits equally. Developers facilitated near-immediate booking of freshly opened slots by building bots that exploit the API. A bot is simply any software application that runs automated tasks over the Internet ~\cite{10.5555/1481092}, to perform simple tasks much faster and more efficiently than a human. In some cases, developers even charge for these services ~\cite{bot_abuse}. As of \formatdate{27}{5}{2021}, there were $1,764$ public Github repositories with code for tools that delivered notifications for available slots or even directly book slots~\cite{bot_abuse} instantly.


Most of these popular bots are programmed to be on online social platforms
such as Telegram or Twitter, which are known to have biased user distributions~\cite{biased_users}. The bots give an unfair advantage to closed circles of people who know that these bots exist. The adverse effects of such bots were felt most in rural areas. The API provides no restriction on the location where a person can book a slot. An individual in an urban district can book slots available in neighbouring rural districts. Technologically literate individuals could use one of the many bots to get instant updates on vaccine availability. In the metropolitan city of Bangalore, the tech-savvy coders used bots to find available centers in the neighbouring rural districts of Chikkaballapur, Ramanagara, and Tumkur~\cite{rural_bot}. A resident in one of these rural districts claimed only about 4-5 locals could get vaccinated on any given day; the rest of the people arrived from Bangalore. 



\subsection{Government steps in to curb the effect of bots}
The government of India took some measures to reduce the effect of disparities that came with the introduction of the CoWIN API (Sec.~\ref{sec:cowin_api_bot}). From \formatdate{3}{6}{2021}, a user was required to enter an OTP (one-time-password) every 15 minutes to continue using the CoWIN app~\cite{security_bots}. On \formatdate{24}{5}{2021}, the government additionally allowed walk-in vaccinations for the 18-44 age group at government-run \COVID vaccinations centers (CVCs)~\cite{instance_1}. The official reason for this (quoted from the press release) - \textit{In case of sessions exclusively organized with online slots, towards the end of the day, some doses may still be left unutilized in case the online appointee beneficiaries do not turn up on day of vaccination due to any reason. In such cases, on-site registration of a few beneficiaries may be necessary to minimize the vaccine wastage.} This policy change benefits people who do not make enough money to afford vaccinations at private hospitals and are not technologically literate to be able to use the CoWIN website or rely on bots for availability information~\cite{instance_1}.


\subsection{Paid access to vaccines}

\label{subsec:21stjune}

From the beginning of the vaccination program (\formatdate{16}{1}{2021}) till \formatdate{30}{4}{2021}, 100\% of the vaccines were procured by the Government of India and provided free of cost to the state governments~\cite{revised_guidelines}, which were administered for free to defined priority groups i.e. healthcare workers and senior citizens. On \formatdate{1}{5}{2021}, the government of India implemented the `Liberalised Pricing and Accelerated National Covid-19 Vaccination Strategy'. Under this scheme, 50\% of the vaccines would be provided for free to the states by the Indian government, but states or private hospitals could buy the other 50\%. A revised version of this policy made the split 75-25 \cite{revised_guidelines}. Note that the private hospitals could charge for the 25\% that was made available to them. The aim was to boost the pace of the vaccination program and incentivise vaccine manufacturers to scale up production. However, many states communicated that the policy negatively impacted the pace of the vaccination program ~\cite{revised_guidelines}. 
Paid access to vaccines is also problematic because of the difference in purchasing power of different demographic groups. This scheme has the potential to aggravate the effect of the wage gap by giving wealthier families an advantage. We further analyse the effectiveness of the June 21 policy across states in Section~\ref{sec:regression_discontinuity}.

\section{Effectiveness of Government Policies}
\label{sec:regression_discontinuity}


To quantitatively assess the effectiveness of government policies on the vaccination drive, we leverage Regression Discontinuity in Time (RDiT). RDiT extends RDD (Regression Discontinuity Design) framework but differs in design as it considers time as a the running variable. This approach imitates quasi-experimental framework and has been used previously in multiple fields such as economics and marketing~\cite{Hausman2017RDiT}. It has even been used to analyse the effect of lockdown policy on \COVID infections in multiple countries~\cite{Shangjun2021RDiT}. We stick to RDiT because other causal inference techniques such as difference in difference framework or propensity score matching was not possible since we did not have any states in control group.


For a given policy to be effective, we should see an increase in the vaccination rates post-implementation of policy compared to pre-implementation of policy. For this analysis, we use the percentage of people that were vaccinated daily as the target response variable. Since this number has a high variance, we smooth it out using a moving average with a window of past 7 days. RDiT is described as follows:
\begin{equation}
\label{eq:1}
y_t  = \beta_0 + \beta_1 t +  \beta_2 \mathbbm{1}(t>0) +  +  \beta_3 \mathbbm{1}(t>0)t
\end{equation}



where $t$ ranges from a negative value to a positive value and $0$ indicating the day of policy implementation. To minimize the effect of unknown confounders~\cite{Hausman2018RegressionDI}, we ensure that $t>0$ does not overlap with any future policies. We also ensure the number of days taken into consideration before and after the policy are equal. $y_t$ represents the statistic we are modelling (the percentage of the population getting vaccinated per day). These models assume that we can approximate the various data points as a straight line defined by $\beta_0$ and $\beta_1$ pre-policy roll-out and $\beta_0+\beta_2$ and $\beta_1+\beta_3$ post-policy roll-out. On applying RDiT, we look at the coefficient associated with the change in membership variable i.e. $\beta_{2}$. We further exclude data from a grace period of one day before and after the policy to account for bursty behaviour.

In particular, we analyse two policies - (1) Government introduced the policy mentioning that 75\% of the vaccines will be provided to the states for free from \formatdate{21}{6}{2021}. (Sec. \ref{subsec:21stjune}) and (2) Policy introduced on \formatdate{4}{6}{2021} through which CoWIN expanded to support 11 other languages which include Hindi, Marathi, Malayalam, Punjabi, Telugu, Gujarati, Assamese, Bengali, Kannada, and Odia ~\cite{CoWIN_language}. The inclusion of the new languages should have helped in making the CoWIN app more accessible. We chose these 2 policies because both of them were nation-wide policies affecting multiple states, and were most isolated, and hence studying their effects was easily possible. We also believe that both of these policies were centered around making vaccination drive more equitable towards marginalized sections of the society. While analyzing the effect of these policies, we were specifically interested in how it affected different sections of the society - hence we focused on the effect of these policies across states and across districts with different rural-ness.


\subsection{June 21st Policy}
On applying RDiT to understand the effect of policy introduced on June 21st, we note that for most of the states the effect of the policy was positive, i.e., it led to more vaccinations. In Figure~\ref{Fig:21stjune_statewise}, all results above the horizontal grey line are significant. The $\beta_{2}$ value indicates the impact of the policy, meaning higher the value, more positive was the impact of the policy on the number of vaccinations. We can see that states like Andaman \& Nicobar Islands and Mizoram benefited the most from this policy, while there was not much improvement in Lakshwadeep and Ladakh.

To understand whether the policy actually assisted in making vaccination drive more equitable, we performed similar analysis on rural and urban districts within each state. We define those districts as rural that have rural ratio in the top 25 percentile, and those as urban, which have rural ratio in bottom 25 percentile. We remove those states which did not have enough number of distinct rural and urban districts from our analysis to avoid including data points with a small sample size. In Figure
3a, most results are significant (above the horizontal grey line). Each state is represented by two dots, blue dot for the rural districts and red dot for the urban districts within that state. In states like Nagaland and Maharashtra, rural districts benefited a lot more. Note that in Arunachal Pradesh, the impact was opposite and urban districts had a more positive impact from this policy, unlike Chattisgarh where the impact was comparable for rural and urban districts.

\subsection{June 4th Policy}
\begin{figure*} \centering
  \begin{subfigure}[b]{0.45\columnwidth}
  \centering
     \includegraphics[width=\linewidth]{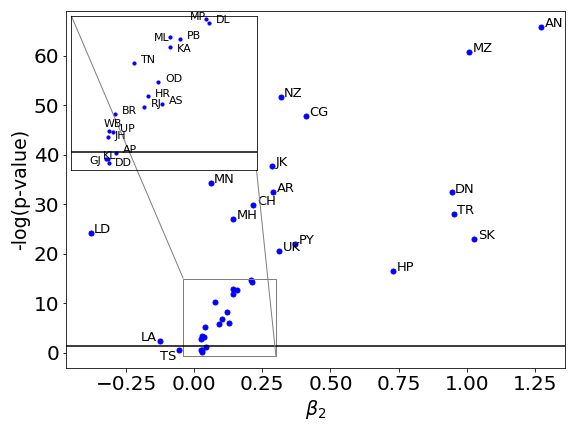}
 \caption{Statewise for 21st June}
 \label{Fig:21stjune_statewise}
  \end{subfigure}
  \begin{subfigure}[b]{0.45\columnwidth}
     \centering
     \includegraphics[width=\linewidth]{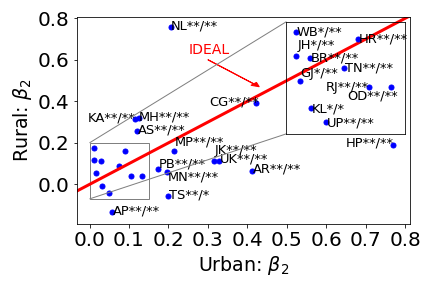}
  \caption{Statewise rural vs urban for 21st June. ** denotes p-value < 0.05, and * denotes p-value > 0.05}
  \label{Fig:21stjune_statewise_ruralurban}
  \end{subfigure}
  \begin{subfigure}[b]{0.45\columnwidth}
  \centering
     \includegraphics[width=\linewidth]{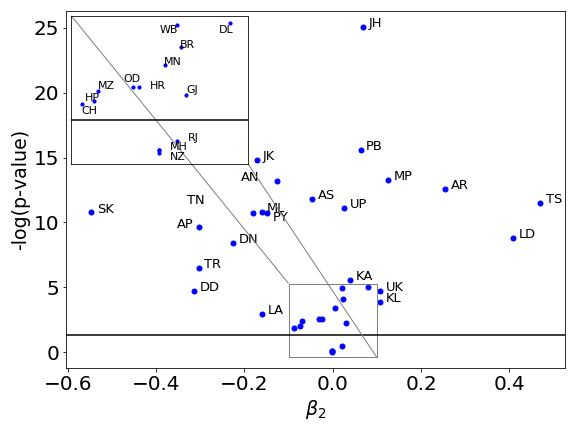}
  \caption{Statewise for 4th June}
 \label{Fig:4thjune_statewise}
  \end{subfigure}
  \begin{subfigure}[b]{0.45\columnwidth}
     \centering
     \includegraphics[width=\linewidth]{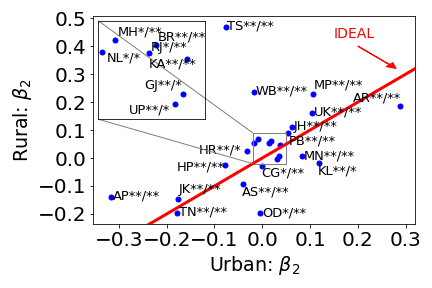}
  \caption{Statewise rural vs urban for 4th June. ** denotes p-value < 0.05, and * denotes p-value > 0.05}
  \label{Fig:4thjune_statewise_ruralurban}
  \end{subfigure}
\caption{Here, the x-axis represents $\beta_{2}$ as explained in eqn~\ref{eq:1}. The y-axis represents the significance of the results, values above black horizontal line are signifcicant. In Fig.~\ref{Fig:21stjune_statewise} and Fig.~\ref{Fig:4thjune_statewise}, we study the effect of the two polices statewise. In Fig.~\ref{Fig:21stjune_statewise_ruralurban} and Fig.~\ref{Fig:4thjune_statewise_ruralurban}, we study the policies' effect on the rural and urban districts of the state. The y-axis represents the top 15 percentile districts (most urban districts) and the x-axis represents the bottom 15 percentile districts (most rural districts). The ideal line depicted shows points where the effect of the policy was equivalent in both groups. Note: Due to clustering of points, a zoomed in version of some areas of the graph is shown.}
  \label{fig:4thjune_regdisc}
\end{figure*}

Similar to the June 21st policy, we also analyze June 4th policy using a similar methodology. We report the effect of the policy to enable more regional languages in the CoWIN app for different states in Figure~\ref{Fig:4thjune_statewise}. The results are mixed in this case. While states like Arunachal Pradesh, Telangana, and Lakshwadeep did have more vaccinations, the number of vaccinations actually went down in Sikkim, Tamil Nadu, and many other states. 

We also check the effect of policy on rural and urban districts. Taking the same definition for binary label of rural and urban districts as above, we report effect of the policy on rural and urban districts in Figure~\ref{Fig:4thjune_statewise_ruralurban}. We note that in some states, one of the data points (either rural or urban) is not significant (below the horizontal line in the graph), so we cannot make conclusions about these states. On the other hand, in states like Telangana, Andhra Pradesh Pradesh, West Bengal, Madhya Pradesh and Jharkhand, rural districts had a significantly more positive impact from the policy. We hypothesize that this could be because the new languages that were introduced were primarily spoken by people of these states.

\section{Data Quality Issues and Observations}
\label{sec:preliminary}


During our study, we found some oddities and inconsistencies that are important to highlight here.  The data from the different sources we leveraged for this study were the only data sources made public and likely motivated several decisions and policies that were made at the state and central level in India. 
For this reason, it is critical that such data is reported accurately to aid effective decision making. We expect all the values reported by the CoWIN dashboard (Sec. \ref{sec:cowin_dashboard_API}) to be consistent since they are the true indicator for the progress of the vaccination drive. However, we found out that this is not the case. In this section, we analyze three such published values.

\label{sec:dash_dis}

\begin{figure*} \centering
  \begin{subfigure}[b]{0.322\columnwidth}
  \centering
     \includegraphics[width=\linewidth]{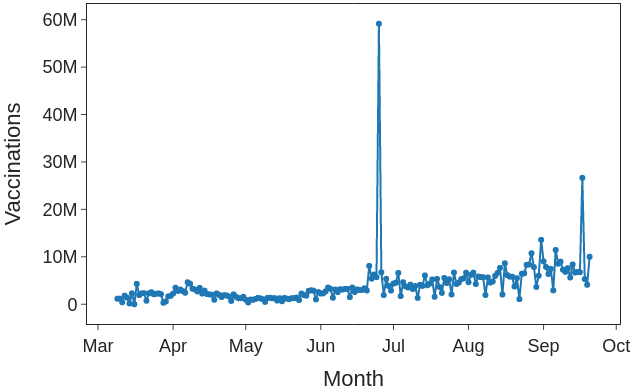}
     \caption{}
     \label{Fig:vaccinations_API}
  \end{subfigure}
  \begin{subfigure}[b]{0.324\columnwidth}
     \centering
     \includegraphics[width=\linewidth]{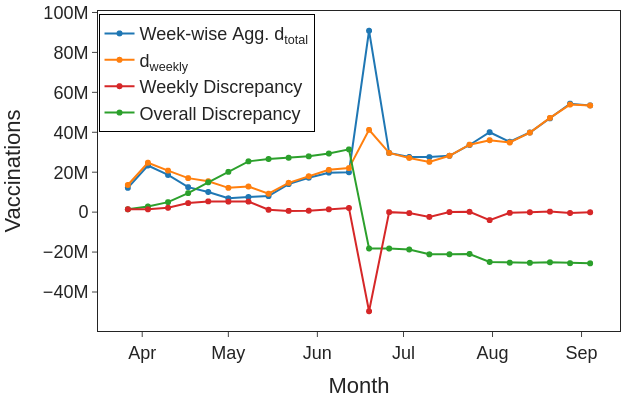}
     \caption{}
     \label{Fig:vaccinations_sub}
  \end{subfigure}
  \begin{subfigure}[b]{0.325\columnwidth}
     \centering
     \includegraphics[width=\linewidth]{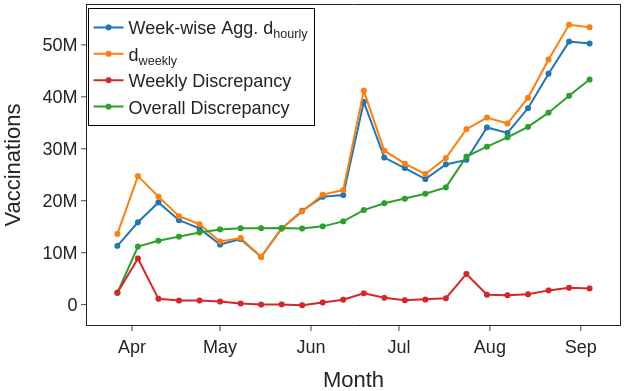}
     \caption{}
     \label{fig:time_API}
  \end{subfigure}
  
\caption{From March 8, 2021 and September 14, 2021, we display (a) Total number of vaccinations based on $d_{total}$. Since the data was cumulative, we performed day-wise subtractions to obtain the number of vaccines administered daily. (b, c) Week-wise and overall cumulative discrepancies between the $d_{weekly}$ data and week-wise aggregated values obtained from $d_{total}$ and $d_{hourly}$. 
}
  \label{Fig:vaccine_field}
\end{figure*}

\begin{enumerate}
    \item \textit{Total till date}: We query the API to get the total number of reported vaccinations until the requested date. We refer to this value as $d_{total}$. In Fig. \ref{Fig:vaccine_field}, we show the distribution of $d_{total}$. Since these values are cumulative (Fig. \ref{Fig:vaccinations_API}), we perform day-wise subtractions to get the graph on the right (Fig. \ref{Fig:vaccinations_sub}), which represents the number of vaccines administered per day based on $d_{total}$.
    \item \textit{Hour-wise vaccinations}: The dashboard depicts the number of vaccinations given hour-wise on any given day, and we record these values from the API as well. We refer to this data as $d_{hourly}$.  We add up the hour-wise values to compute the total number of vaccinations on that date. The number of vaccinations administered day-wise based on $d_{hourly}$ is displayed in Fig. \ref{fig:time_API}. Notice how Fig. \ref{fig:time_API} and Fig. \ref{Fig:vaccinations_sub} depict significantly different distributions for the same statistic.
    \item \textit{Week-wise vaccinations}: We obtain week-wise vaccination statistics, i.e., the total number of vaccine doses administered each week for most of 2021. We refer to this distribution as $d_{weekly}$.
\end{enumerate}

\label{sec:vaccination_field}

To compare these values, we consider $d_{weekly}$ to be the baseline and aggregate data from $d_{total}$ and $d_{hourly}$ to obtain week-wise statistics. Fig. \ref{Fig:vaccinations_sub} demonstrates how the week-wise aggregation based on $d_{total}$ compares with $d_{weekly}$. The most notable mismatch is in the week of June 19 to June 25. According to the API, 90.84 million vaccinations were given that week. However, the dashboard displays 41.20 million. That is a difference of precisely 49,642,682 vaccine doses just for that week. The cumulative difference (shown as \textit{overall discrepancy} in the graph) between the statistic derived from $d_{total}$ and the $d_{weekly}$ values also takes a sharply negative turn in that week. Note that this is the week when the government started buying 75\% of the vaccines as per the `Revised Guidelines for implementation of National Covid-19 Vaccination Program' hospitals ~\cite{revised_guidelines} (discussed in Section \ref{subsec:21stjune}).


We aggregate the $d_{hourly}$ values week-wise and compare them with $d_{weekly}$ in Fig. \ref{fig:time_API}. We can observe that the aggregation closely follows $d_{weekly}$, unlike the previous comparison. Note how the cumulative difference (shown as \textit{overall discrepancy} in the graph) stays close to zero for most points as one would expect. There are some discrepancies nonetheless. In this case, the highest week-wise discrepancy is 8,891,282 vaccinations in the week of April 3-9.  We obtained similar results for some other statistics for age-wise data as well. We also noted that different end-points of the API returns different values. Moreover, these values are not consistent with the numbers being displayed on the CoWIN dashboard. After carefully studying this data source, we carried out our analyses (in Sec. \ref{sec:regression_discontinuity}) using $d_{hourly}$ as it most closely matches $d_{weekly}$ (Fig. \ref{fig:time_API}), and is the most granular. However, these inconsistencies raise several questions -- what are the accurate numbers? Which data should analysts use to perform analysis? What published statistics should policy makers rely on to make decisions?

\section{Implications}






The effect of the inequities that arose during the pandemic are well known. Some sections of the population suffered more, likely due to the distribution of class economics in society. For example, people who have to go out for their jobs and cannot maintain social distancing have been very vulnerable to \COVID exposure. Vaccination-related policies are further extenuating such inequities. In this paper, we reviewed India's vaccination policies and potential inequities that have been introduced or further aggravated. We also identified inconsistencies in the data reported about the vaccination drive. We believe that our work has multiple real-world implications, which we list below.

\noindent
\textbf{Guidance for future policies:}  Analyzing existing policies from an equity perspective can lead to greater understanding and provide guidance on designing and developing future vaccination-related policies such that they are more equitable. This knowledge pertaining to fairness can be applied onto other forms of government planning, especially in health, and optimum resource allocation processes. Although minimal in India, large amount of work has been done elsewhere in helping government policymakers to plan, allocate, and utilize available resources~\cite{eviction_mitigation, Colleen2021bailout}. Our findings on the effectiveness of policies in India can be used to come up with checks to enforce that future policies are as equitable as possible for all sections of society.

\noindent    
\textbf{Importance of data to understand policy effect:} We advocate analysing the effectiveness and outcome of implemented policies by using data-based analyses like the ones presented in this paper. Quantitative study  makes it possible to take more informed decisions and make appropriate reforms in policy implementation. As more and more government policies across the globe are formulated and evaluated using data-driven inferences, creating credible data which can aid policymaking is becoming an important problem to address ~\cite{Rukmini2021}. Several previous works, especially in the US, have exploited access to data to analyse policies~\cite{Colleen2021bailout, GLOVER202035}. We thus highlight the need for granular and diverse data to conduct such studies.

\noindent
\textbf{Data transparency and quality:} We demonstrate accuracy and quality concerns in the vaccination data published by the government and highlight the need to resolve the same. Such inconsistencies must be resolved as reliable data is necessary to back economic policies and research~\cite{Blumenstock2015}. There are national forums already like National Data Quality Forum (NDQF)~\cite{ndqf}, and these problems should be included in their mandate. There are multiple sub-populations in India that do not have access to technical solutions that are developed and deployed in India because of financial status, lack of exposure to technology, rural/urban settings, etc. To realise the true potential of India, breaking these barriers via transparent and data-driven policymaking will be very critical ~\cite{Karnik2021}. Further, fairness, accountability, and transparency have to be key pillars of that transition.

\section{Limitations}

Our study, like any data-based study on real-world phenomena, has limitations. In this section, we try our best to enumerate them.

\noindent
\textbf{Data Issues}:
The most reliable source to obtain data on demographics at an all-India level is the census conducted by the government. However, India's most recent census was conducted in 2011. This makes our data related to the distribution of rural and urban populations outdated. Census in India was planned for 2021, but it was not conducted due to many reasons~\cite{indiaTV2021census}. 

\noindent
\textbf{Unreliability of Vaccination Data:}
As described in Section~\ref{sec:preliminary}, the government-reported data on vaccinations is inconsistent and unreliable. This irregularity hurts our research as well and we couldn't make conclusive claims in some of our analyses. This observation also emphasizes our ask for more open data to quantitatively analyze the effects of policies. 

\noindent    
\textbf{Inference Issues:}    
In Sections~\ref{sec:correlation} and \ref{sec:regression_discontinuity}, we use features to test the bias and assess how rural ratio affected vaccine distribution and accessibility. We try our best to isolate the effects and make sure to test the model's assumptions. However, there may be some unmeasured or latent confounding variables that could be driving the distribution of vaccines, which we might not have considered.

\noindent
\textbf{Unknown Requirements and Constraints:}
We acknowledge that we are conducting our study from a perspective outside the government, and we may not have access to certain information or restrictions the government is accounting for when making policies and decisions.

\section{Discussion and Conclusion}


The COVID-19 pandemic has affected some people much worse than others; vaccinating the population is crucial to combat the effects of the pandemic. It is essential that the distribution of vaccines is done equitably. Throughout the pandemic, technology, data, and policy-making seemed to have played a big role, especially in India. We found various sources of data being published regarding the number of cases, availability of vaccines, phone applications and bots being developed, distribution of vaccines, and planning and execution of policies at the health centre level. We used data reported about vaccination statistics to analyse the inequities. We found that some states were more biased in their distributions. We analysed the effectiveness of two major policies that were introduced, and demonstrated how some of these policies failed at achieving equitable distribution in some states. It is also essential that the vaccination data made available is consistent and accurate to inform reliable policies and decisions. We highlight inconsistencies in the vaccination data made available on the COWIN Dashboard. Through quantitative analysis, we point out critical inequities in the administration of vaccinations and advocate that future policies be made by taking equity and transparency into account.


\newpage
\bibliographystyle{ACM-Reference-Format}
\bibliography{sample-base}


\end{document}